\documentclass[a4paper,12pt,oneside]{article}
\usepackage[koi8-r]{inputenc}
\usepackage[english]{babel}
\usepackage{graphicx}
\usepackage{amsmath}
\usepackage{amssymb}
\usepackage{euscript}
\begin{document}

\title{On the nature of veiling of classical T Tauri stars spectra in the
near-IR spectral band.}

\author{A.V.Dodin, S.A.Lamzin }

\date{ \it \small
Sternberg Astronomical Institute of Lomonosov Moscow State University,
Universitetskij pr. 13, Moscow, 119992 Russia
\footnote {Send offprint requests to: A. Dodin e-mail: dodin\_nv@mail.ru}
}

\maketitle

\medskip



\medskip


\section*{Abstract}

  It is shown that the existence of a hot accretion spot on the surface of
classical T Tauri stars allows to explain observed veiling of their
photospheric spectrum not only in the visible but also in the near infrared
spectral band.


\section*{Introduction}

 Classical T Tauri stars (CTTS) are young ($t<10^{7}$ yr), low mass
$(M\leqslant 3\,M_\odot)$ stars at the stage of gravitational contraction
towards the main-sequence, activity of which is caused by magnetospheric
accretion of protoplanetary disc matter.  It has long been known that the
depths and equivalent widths of photospheric lines in the optical and
ultraviolet spectra of T Tauri stars are smaller than that of main-sequence
stars of the same spectral types.  Recently this effect has been found in
the near-infrared (NIR) spectral band -- see Fischer et al.  (2011) and
references therein.

  It is widely accepted that the veiling is caused by an additional
continuous emission, which is formed in accretion hot spot on stellar 
surface due to heating of CTTS's atmosphere by X-ray and EUV radiation of
accretion shock.  However Fischer et al.  (2011) found that commonly
used hot spot models of Calvet \& Gullbring (1998), which assumed that the
spot radiates in continuum only, cannot simultaneously explain the veiling
at visible and NIR wavelengths because spot's continuum emission flux
dereases too rapidly if wavelength increasing.  According to Muzerolle et
al.  (2003) emission of dust component of protoplanetary disk gives
noticeable excess relative to the photospheric radiation only at
wavelengths $\lambda > 2$ $\mu$m, so to explain observed veiling 
near 1~$\mu$m Fischer et al.  (2011) have suggested that there is an
additional source of the continuum emission with temperature 2500-5000~K.

   Gahm et al. (2008) found from the analysis of highly veiled CTTS's
spectra that emission lines, which are formed in the accretion spot along
with continuum, contribute significantly to decreasing of photospheric
line's depth.  Dodin \& Lamzin (2012) confirmed this discovery by
theoretical modeling and also demonstrated that "veiling by lines"\, is
important for both CTTS with high and small veiled spectra.  Moreover they
found that at given effective temperature of the star $T_{ef}$ the relative
contribution of lines to veiling increases with decreasing of accretion
flux, which is defined by the relation $F_{ac}=\rho_0 V_0^3/2,$ where
$\rho_0$ and $V_0$ are pre-shock gas density and velocity respsectively.

  Intensity of hot spot's continuum emission in Dodin \& Lamzin (2012) model
decreases when one moves from visible to NIR spectral band as well as in the
models of Calvet \& Gullbring (1998).  However wavelenght dependence of
veiling by lines should be different, because it depends on gas temperature
and density distribution in the stellar atmosphere heated by accretion shock
radiation, as well as on parameters of individual lines, rather than on hot
spot's effective temperature. As the result dependence of the total
(lines+continuum) veiling on wavelength in optical band becomes
non-monotonic instead of monotonic in the case of veiling by continuum only
-- see Fig.\,7 in Dodin \& Lamzin (2012).

  If near-IR veiling is caused mostly by lines then it would allow to
explain observed CTTS's veiling at wavelengths near 1~$\mu$m, not involving
an additional source of continuum emission.  The aim of our work is to
test this hypothesis.


\section*{The dependence of veiling on $\lambda$ at high and low accretion
fluxes in the frame of homogeneous spot model.}

  We will consider in this paper spectra of the system "star+round
homogeneous spot"{}, calculated by methods, described by Dodin \&
Lamzin(2012).  The only difference is that we consider here not only visible
but also NIR spectral band, more precisely from 0.45 to 1.2~$\mu$m, i.e.  up
to maximum wavelength for which atomic data for spectral lines are aviable
in the ATLAS package that we use.  The expression "homogeneous spot"{} means
that $V_0$ and $\rho_0$ values are assumed to be constant across the
accretion flow.  In what follows we will use pre-shock gas number density
rather than gas density: $N_0\approx\rho_0/2.2\times10^{-24}.$

  Remind that it is used to characterize the veiling by the quantity
$$
r_\lambda = {EW_0 \over EW}-1,
$$
where $EW$ and $EW_0$ are equivalent widths of some photospheric line in
spectra of CTTS and template star of the same spectral type respectively. 
To characterize veiling in some spectral region one presents
$r_\lambda$-value avaraged over all absorption lines of this band.

%
\begin{table}[h!]
  \caption{List of photospheric lines used to calculate $r_\lambda$}
 \label{highveiled}
\begin{center}
\begin{tabular}{|c c|c c|c c|}
\hline
                  & $\lambda$, {\AA} &         & $\lambda$, {\AA}&         & $\lambda$, {\AA}  \\
\hline
 Ti\,I, Mg\,I     &     4783.3     &      Fe\,I            &     6173.3     &      Fe\,I     &     9173.2\\
 Ni\,I, Cr\,I     &     4829.2     &      Cr\,I            &     6330.1     &      Ti\,I     &     9599.6\\
 Fe\,I            &     4903.3     &      Fe\,I, Fe\,I     &     6400.1     &      Ti\,I     &     9832.1\\
 Cr\,I, Fe\,I     &     4942.5     &      Ca\,I            &     6462.6     &      Fe\,I     &     9889.0\\
 Ti\,I, Fe\,I     &     4991.1     &      Ni\,I            &     6643.6     &      Ti\,I     &     10003\\
 Fe\,I, Fe\,I     &     5027.2     &      Ti\,I            &     6743.1     &      Fe\,I     &     10155\\
 Fe\,I, Cr\,I     &     5139.5     &      Fe\,I            &     6806.8     &      Fe\,I     &     10167\\
 Fe\,I, Fe\,I     &     5273.2     &      Al\,I, Co\,I     &     7084.9     &      Fe\,I     &     10341\\
 Fe\,I            &     5391.5     &      Fe\,I            &     7292.8     &      Cr\,I     &     10486\\
 Fe\,I            &     5476.6     &      Ca\,I            &     7326.2     &      Fe\,I     &     10532\\
 Ca\,I            &     5598.5     &      Ti\,I, Fe\,I     &     7440.8     &      Ti\,I     &     10662\\
 Fe\,I, Cr\,I     &     5682.5     &      Fe\,I            &     7583.8     &      Ti\,I     &     10677\\
 Fe\,I, Fe\,I     &     5762.7     &      Ni\,I            &     7727.6     &      Ti\,I     &     10775\\
 Cr\,I, Fe\,I     &     5791.0     &      Ti\,I, Ti\,I     &     7978.8     &      Cr\,I     &     11157\\
 Ca\,I            &     5857.5     &      Fe\,I            &     8075.2     &      Cr\,I     &     11339\\
 Fe\,I            &     5916.3     &      Mg\,I, Fe\,I     &     8310.9     &      Cr\,I     &     11398\\
 Fe\,I, Ti\,I     &     5952.9     &      Fe\,I            &     8616.3     &      Fe\,I     &     11422\\
 Fe\,I            &     6027.1     &      Cr\,I            &     8976.9     &      Fe\,I     &     11595\\
 Ti\,I            &     6085.2     &      Fe\,I            &     9070.4     &      Ti\,I     &     11797\\
 Ca\,I            &     6122.2     &      Ca\,I            &     9099.1     &      Ca\,I     &     11956\\
\hline
 \multicolumn{6}{p{12cm}}
{\footnotesize {\bf Note.}  Non-LTE cflculations were carried out for 
Ca\,I lines (Dodin et al., 2013).}
 \end{tabular}
  \end{center}
\end{table}
%

  We will demonstrate the contribution of emission lines to the total
veiling on the example of two models: with relatively high and relatively
low accretion flux $F_{ac}.$ These models reproduce (see Dodin et al., 2013
for details) 0.47-0.80~$\mu$m spectra of CW Tau and BP Tau with reasonable
accuracy, such as it is the same spectra that Fischer et al.  (2011) used
for optical veiling measurements.  They were observed on 2006 November 30
and were taken from Keck Observatory Archive
http://www2.keck.hawaii.edu/koa/public/koa.php.

  In the case of CW Tau the parameters of the model were the following: the
effective temperature and the gravity of the star were $T_{ef}=4750$~K and
$\log g = 4.0$, respectively; $N_0=10^{12}$ cm$^{-3}$, $V_0=350$
km\,s$^{-1}$, the relative area of the spot $f=0.20,$ the angle between
spot's symmetry axis and the line of sight $\alpha=0.$ This model is the
example of a large spot (it occupies 40\,\% of {\it visible} stellar
hemisphere) with relatively low infall gas density $N_0$ and low accretion
flux $\log F_{ac}\approx 10.7.$

  We will see below (see also Dodin \& Lamzin, 2012) that $r_\lambda$
strongly varies from one line to another, and therefore lines, used for
veiling measurements, should be always specified.  Unfortunately until now
nobody does it, interpreting the scatter of $r_\lambda$-values of individual
lines in the considered spectral band as a "random error"\, of measurements. 
We have choosen for our veiling measurements deep absorption lines (without
any signs of emission core) from our model spectrum which practically coincides
with observed one.  The list of these lines is presented in the Table.

\begin{figure}
 \begin{center}
\includegraphics[scale=0.5]{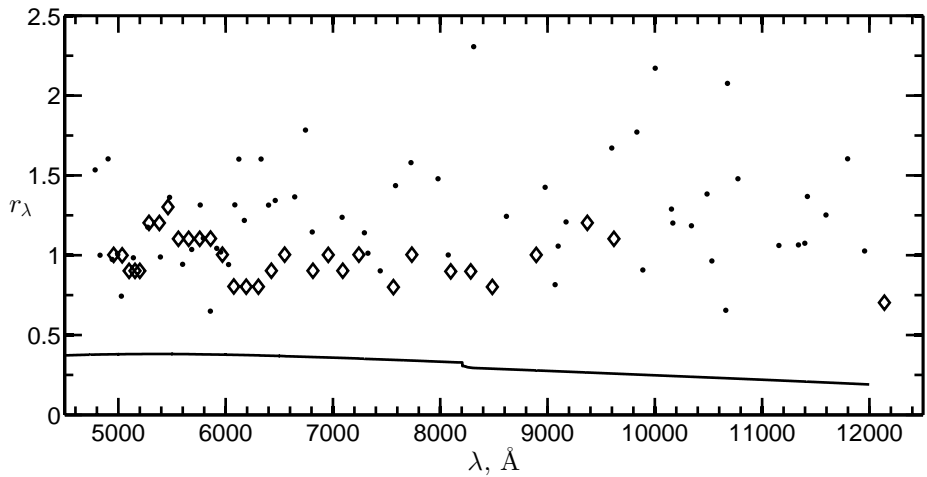}
\caption{The veiling as function of wavelength in the model with the 
following parameters: $T_{ef}=4750$~K, $\log g = 4.0,$ $N_0=10^{12}$
cm$^{-1}$, $V_0=350$ km\,s$^{-1}$, $f=0.20$, $\alpha=0^o.$ The dots indicate
the total veiling in the lines from the Table, while the solid line
indicates the veiling due to the hot continuum only. The diamonds represent
veiling measurements of CW Tau spectrum by Fisher et al. (2011).  
}
  \label{lowfacc}
 \end{center}
\end{figure}

  The dependence $r_\lambda (\lambda)$ that we calculated from CW Tau model
spectrum is shown on Fig.\,\ref{lowfacc}.  As far as theoretical and
observed spectra coincide in the optical band our $r_\lambda$-values in this
range should be identical to that of Fischer et al.  (2011) up to the choice
of spectral lines used for veiling measurements, and they indeed do as can
be seen from the figure. At the same time the NIR veiling {\it predicted}
by the model is also practically coinsides with Fischer et al. data, who have
measured the NIR veiling from the IR spectrum, observed simultaneously with
the optical one.

%
%

  In addition to the total (line+continuum) veiling of the individual lines
we plotted on the figure the curve, which indicates the veiling of
photospheric lines caused by spot's continuum emission only.  It can be seen
that in the considered case contribution of lines in decreasing of depth of
photospheric lines is much larger than that of emission continuum, and it
explains why values of $r_\lambda$ at 5000 {\AA} and at 1~$\mu$m are
practically the same in CW Tau spectrum.

   We would like to pay attention to the important feature of the veiling by
lines, which in fact consists of two components: 1) superposition of
emission lines onto respective absorption lines of CTTS's photosphere; 2)
decreasing of surface of undisturbed photosphere due to the presence of
a spot (cold or hot does not matter) or even due to eclipse of some part of
stellar disk, for example by opaque dust cloud.  The second effect is
caused by the fact that the observed flux is the result of integration of
radiation specific intensity over visible stellar hemisphere.

  The removal of any part of the stellar surface from the integration region
leads not only to decreasing of observed flux, but also to deformation of
line's profiles and to changing of their equivalent width due to differences
in the limb darkening low in lines and continuum.  For example we found that
if to equate the intensity of spot's radiation to zero in the CW Tau model,
i.e.  to replace the hot spot by the absolutely cold one, then the veiling
for various lines will vary chaotically from -0.15 to +0.05 in the blue and
from -0.05 to +0.25 in the IR spectral bands.  Thus the radiations from the
spot and the star should be summed up accurately, especially in the case of
low veiling or in the case of possible presence of large cold or hot spots.

%
\begin{figure}
 \begin{center}
  \includegraphics[scale=0.5]{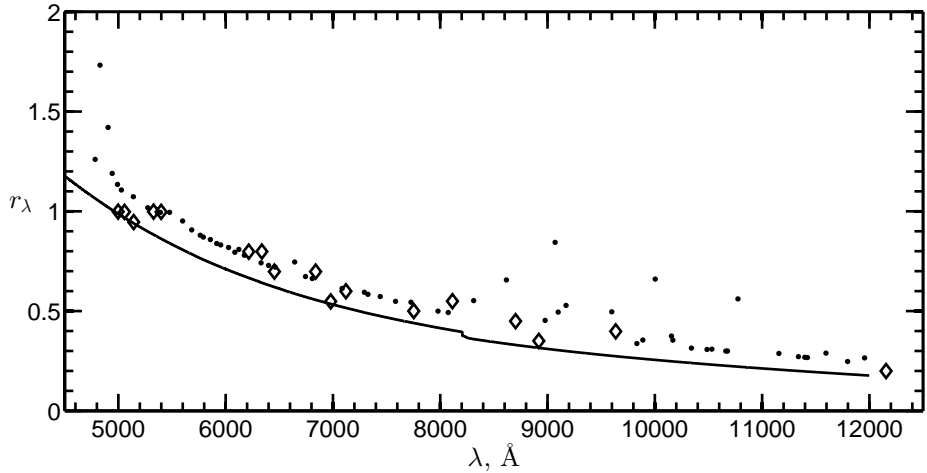}
 \caption{ The veiling as function of wavelength in the model with the
following parameters: $T_{ef}=4000$~K, $\log g = 3.5,$ $N_0= 10^{13}$
cm$^{-1}$, $V_0=350$ km\,s$^{-1}$, $f=0.01$, $\alpha=0^o.$ The dots indicate
the total veiling of the lines from the Table, while the solid line
indicates the veiling due to the hot continuum only. The diamonds represent
veiling in BP Tau spectrum measured by Fisher et al. (2011).
}
  \label{highfacc}
 \end{center}
\end{figure}
%

  As the second example consider the model with larger $F_{ac}$ value, which
well reproduces the visual spectrum of BP Tau.  The parameters of the model
are the following: $T_{ef}=4000$~K, $\log g=3.5,$ $N_0=10^{13}$ cm$^{-3}$,
$V_0=350$ km\,s$^{-1}$, $f=0.01,$ $\alpha=0^o.$ This model is the example of
a small spot (it occupies only 2\,\% of the {\it visible} stellar
hemisphere), but with large pre-shock gas density $N_0$ and large accretion
flux $\log F_{ac}\approx 11.7.$

  It can be seen from Fig.\,\ref{highfacc} that the model well reproduces
the observed veiling in the range from 0.47 to 1.2~$\mu$m.  Note that the IR
spectrum of BP Tau, in which Fischer et al.  have measured the veiling, was
observed simultaneously with the optical one.  In the case of BP Tau the
veiling is caused predominantly by continuum emission, and therefore
$r_\lambda$ almost monotonically decreases with wavelength.  It would seem
that Calvet \& Gullbring (1998) model, which assumes that hot spot radiates
in continuum only, could explain BP Tau observations, and therefore it is
unclear why Fischer et al.  (2011) who used this model, could not agree
theory with observations.


\section*{Conclusions}

   We have shown that if to calculate accretion hot spot spectrum taking
into account not only continuum but also line emission and properly sum up
contributions of the spot and undisturbed stellar photosphere then the
observed veiling of CTTS spectra in the near-IR band can be explained
without additional sources of veiling continuum.

  Fischer et al. (2011) found that the shapes of $r_\lambda(\lambda)$
dependences in CTTS spectra can be very different.  It follows from our
consideration of homogeneous accretion spot models that in the case of a
large spot with low $F_{ac}$ an average value of $r_\lambda$ is almost
constant from 0.5 to 1.2~$\mu$m, and in the case of a small spot with high
$F_{ac}$ the veiling drops rapidly with wavelength increasing.  It means
that models with inhomogenious distribution of $N_0$ and $V_0$ parameters
across the spot can produce various shapes of $r_\lambda(\lambda)$ curves:
from the almost flat veiling up to the veiling, which concentrated almost
complitely in UV and visual spectral bands.  Note that the real spots should
certainly be inhomogeneous (Romanova et al., 2004).

  We would like to note that neglecting with contribution of lines in veiling
should lead to an error in estimation of interstellar extinction.  Indeed to
compensate hypothetical additional continuum emission it is necessary to
increase its extinction.  Apparently it is why Fisher et al.  (2011) found
that the extinction determined from IR spectra is systematically larger than
that determined from optical ones.

\bigskip

  We thank Dr. L.Ingleby and her colleagues, who drew your attention to
the problem of near-IR veiling.  This research has made use of the Keck
Observatory Archive, which is operated by the W.M.  Keck Observatory and the
NASA Exoplanet Science Institute (NExScI), under contract with the 
the National Aeronautics and Space Administration.  S.Dahm is the principal
investigator of BP Tau and CW Tau observations, used in this paper.
The work was supported by the Program for Support of Leading Russian
Scientific Schools (NSh-5440.2012.2).


\end{document}